\documentclass[oldversion,twocolumn]{aa}
\usepackage{natbib}
\usepackage{epsfig}
\usepackage{txfonts}

\def\ms{\,m\,s$^{-1}$}         
\def\msol{M_\odot}             
\def\mjup{M$_{Jup}$}             

\begin{document}

\title{ The HARPS search for southern extra-solar planets\thanks{Based
on observations made with the HARPS instrument on the ESO 3.6 m
telescope at La Silla Observatory under programme ID 072.C-0488(E).}
}

\subtitle{XVII. Six long-period giant planets around BD\,-17 0063, HD\,20868, HD\,73267, HD\,131664, HD\,145377, HD\,153950}

\author{
C. Moutou\inst{1}
\and M. Mayor\inst{2}
\and G. Lo Curto\inst{3}
\and S. Udry\inst{2}
\and F. Bouchy\inst{4}
\and W. Benz\inst{5}
\and C. Lovis\inst{2}
\and D. Naef\inst{3}
\and F. Pepe\inst{2}
\and D. Queloz\inst{2}
\and N.C. Santos\inst{6}
}

\offprints{C. Moutou}

\institute{
Laboratoire d'Astrophysique de Marseille, OAMP, Universit\'e Aix-Marseille \&
CNRS, 38 rue Fr\'ed\'eric Joliot-Curie, 13388 Marseille cedex 13, France\\
\email{Claire.Moutou@oamp.fr}
\and 
Observatoire de Gen\`eve, Universit\'e de Gen\`eve, 51 ch.des Maillettes, 1290 Sauverny, Switzerland.
\and
ESO, Alonso de Cordoba 3107, Vitacura Casilla 19001, Santiago, Chile
\and
Institut d'Astrophysique de Paris, 98bis bd Arago, 75014 Paris, France
\and
Physikalisches Institut Universit\"at Bern, Sidlerstrasse 5, 3012
Bern, Switzerland
\and
    Centro de Astrof{\'\i}sica, Universidade do Porto, Rua das Estrelas,
    4150-762 Porto, Portugal
}

\date{Received ; accepted }

\abstract{We report the discovery of six new substellar companions of
  main-sequence stars, detected through multiple
  Doppler measurements with the instrument HARPS installed on the ESO
  3.6m telescope, La Silla, Chile. These extrasolar planets are orbiting the
  stars BD\,-17 0063, HD\,20868, HD\,73267, HD\,131664, HD\,145377, HD\,153950. The orbital
  characteristics which best fit the observed data are depicted in this paper,
  as well as the stellar and planetary parameters. Masses of the companions
  range from 2 to 18 Jupiter masses, and periods range from 100 to 2000 days.
  The observational data are carefully analysed for
  activity-induced effects and we conclude on the reliability of the observed
  radial-velocity variations as of exoplanetary origin. Of particular interest
  is the very massive planet (or brown-dwarf companion) around the metal-rich HD\,131664 with
  $m_2 \sin{i}=$ 18.15 \mjup\, and a 5.34-year orbital period. These new
  discoveries reinforces the observed statistical properties of the exoplanet
  sample as known so far.
}

\maketitle

\section{Introduction}

The HARPS \footnote{High-Accuracy Radial-velocity Planet Searcher}
instrument \citep{pepe03,messenger03} has been in operation since October 2003 on the 3.6m
telescope in La Silla Observatory, ESO, Chile. It has allowed so far the
discovery of several tens of extrasolar systems, among which very low-mass
companions (e.g., \citet{mayor08}). The strategy of HARPS observations
inside the Guaranteed Time Observation program is adapted to different
target samples. High-precision is achieved on a sub-sample
of bright stars, known to be stable at a high level. In addition, a larger,
volume-limited sample of stars are being explored at a moderate precision 
(better than 3 \ms or signal-to-noise ratio of 40) in order to complete our view
of exoplanets' properties with extended statistics. The HARPS sample completes
the CORALIE sample with stars from 50 to 57.5 pc distance, and together, these
samples contain about 2500 stars. The results presented in this
paper concern this wide exploratory program at moderate precision. Earlier
findings in this stellar sample consist in 8 giant planets, which have been
presented in \citet{pepe04}, \citet{moutou05}, \citet{locurto06}, and \citet{naef07}.
The statistical properties of these planets encounter those described in the
literature \citep{marcy2005, udry2007}, regarding the frequency of planets and
the distribution of their parameters.

We report the discovery of six new planets in the volume-limited sample of
main-sequence stars, using 
multiple HARPS Doppler measurements over 3 to 5 years. They are
massive and long-period planets. 
Section \ref{star} describes the characteristics of the parent stars, and 
Section \ref{planet} presents the Doppler measurements and discusses the planetary orbital solutions. 

\section{Characteristics of the host stars}
\label{star}

The host stars discussed here are: BD\,-17 0063, HD\,20868, HD\,73267,
HD\,131664, HD\,145377, and HD\,153950. We used the $V$ magnitude and $B-V$ color
index given in the {\sc Hipparcos} catalog \citep{hip}, and the Hipparcos parallaxes $\pi$
as recently
reviewed by \citet{vanleeuwen07}, to estimate the absolute magnitude $M_V$. The bolometric correction of \citet{flower96}
is then applied to recover the absolute luminosity of the stars. 

Spectroscopic parameters $T_{eff}$, log$g$, and $[Fe/H]$ were derived from a set of FeI and FeII lines
\citep{santos2004} for which equivalent widths were derived with ARES (Automatic Routine for line
Equivalent width in stellar Spectra, \citet{sousa07,sousa08}) on the HARPS spectra.
The error-bars reflect the large number of
FeI and FeII lines used, and a good precision is obtained, especially on the
effective temperature. For gravity and metallicity estimates we are limited by
systematics, which are included in the error bars. 

We finally estimate the stellar mass and age, from $T_{eff}$, $[Fe/H]$, and parallax estimates, by
using Padova models of \citet{girardi00} and its web interface as described in
\citet{dasilva06}. Errors are propagated and estimated using the Bayesian method.
The stellar radius is finally estimated from the 
simple relationship between luminosity, temperature and radius. 

From the HARPS cross-correlation function, we may derive an estimate of
the projected rotational velocity of the star, $v$sin$i$. The measurement of
the core reversal in the calcium H \& K lines provides an estimate of the
chromospheric activity log$R'_{HK}$ (see method in \citet{santos00}). The
error bars of this quantity include the scatter as well as some systematics;
these are particularly
large for the faintest and coolest stars, with typical signal-to-noise ratio
of 10 in the region of the calcium doublet.
All mesured stellar parameters and their errors are given in Table 1. A short presentation of the host
stars follows.

\subsection{BD\,-17 0063, HD\,20868, and HD\,73267: three quiet K-G stars}
BD\,-17 0063 is a main-sequence K5 and HD\,20868 is slightly evolved 
K3/4 IV star. Both are slow rotators which do not
exhibit a significant activity jitter. They have a metallicity compatible with
the Sun metallicity and masses of respectively 0.74 and 0.78 $\msol$ and
estimated age of more than 4 Gyr.
HD\,73267 is somewhat more massive with 0.89 $\msol$. It also shows no significant
activity, rotates slowly and has a solar metallicity. It is a 7 Gyr-old G5
dwarf. 
{\bf The rotation period of our sample
stars can be extrapolated from the activity level using relations in \citet{noyes84} (colour to convection turnover time relations) and \citet{mamajek2008} (Rossby number to  log$R'_{HK}$ relations). We find rotation periods of 39, 51 and 42 days, respectively for  BD\,-17 0063, HD\,20868, and HD\,73267.}

\subsection{HD\,131664, HD\,145377, and HD\,153950: two early G and one late F
  dwarf stars}
The other three stars HD\,131664, HD\,145377, and HD\,153950 are slightly more massive
than the Sun with respecively 1.10, 1.12 and 1.12 $\msol$. HD\,145377 is the most
active one and its age is estimated around 1 Gyr. Their rotational periods are
also shorter than for the first group of lower mass stars, {\bf with 22, 12 and 14 days respectively for HD\,131664, HD\,145377, and HD\,153950}. HD\,131664 and HD\,145377 are both metal-rich stars with $[Fe/H] =$ 0.32 and 0.12,
respectively, while  HD\,153950 has a metallicity close to solar.

\begin{table*}
\caption{Observed and inferred stellar parameters for the planet-hosting stars presented in this paper.{\bf $^a$ The rotation periods of the stars are not derived from observations, but indirectly inferred from $\log R'_{\mathrm{HK}}$ with relationships in \citet{mamajek2008} and \citet{noyes84}.}}
\label{TableStars}
\centering
\begin{tabular}{l l c c c c c c}
\hline\hline
\multicolumn{2}{l}{\bf Parameter} & \bf BD-17 0063 & \bf HD\,20868 & \bf HD\,73267 & \bf HD\,131664 & \bf HD\,145377 & \bf HD\,153950\\
\hline 
Sp &  & K5V & K3/4IV & G5V & G3V & G3V & F8V\\
$V$ & [mag] & 9.62 &9.92 & 8.90 & 8.13 & 8.12 & 7.39\\
$B-V$ & [mag] &  1.128 & 1.037 & 0.806 & 0.667 & 0.623 & 0.565\\
$\pi$ & [mas] & 28.91 (1.27) & 20.42 (1.38) & 18.21 (0.93) & 18.04 (0.73) &
18.27 (0.94) & 20.16 (0.70)\\
$d$ & [pc] & 34.6 (1.5) & 48.9 (3.5) & 54.91 (3.0) & 55.43 (2.3) & 57.7 (3.0) & 49.6 (1.8)\\
$M_V$ & [mag] & 6.92 & 6.47 & 5.20 & 4.41 & 4.31 & 3.91\\
$B.C.$ & [mag] & -0.49 & -0.41 & -0.195 & -0.08 & -0.055 & -0.038\\ 
$L$ & [$L_{\odot}$] & 0.21 (0.02) & 0.296 (0.04) & 0.783 (0.09) & 1.46 (0.13)
& 1.56 (0.17) & 2.22 (0.17)\\
$T_{\mathrm{eff}}$ & [K]   & 4714 (93) &4795 (124) & 5317 (34) & 5886 (21) &
6046 (15) & 6076 (13)\\
log $g$            & [cgs] & 4.26 (0.24) & 4.22 (0.26) & 4.28 (0.1) & 4.44
(0.1) & 4.49 (0.1) & 4.37 (0.1)\\
$\mathrm{[Fe/H]}$  & [dex] & -0.03 (0.06) &0.04 (0.1) & 0.03 (0.02) & 0.32
(0.02) & 0.12 (0.01) & -0.01 (0.01) \\
$M_*$ & [M$_{\odot}$]  & 0.74 (0.03) & 0.78 (0.03) & 0.89  (0.03) & 1.10
(0.03) & 1.12 (0.03) & 1.12 (0.03)\\
$v\sin{i}$ & [km s$^{-1}$] & 1.5 &1.1 & 1.65 &  2.9 & 3.85 & 3.0 \\
$\log R'_{\mathrm{HK}}$ &  & -4.79 (0.1) & -4.99 (0.1) & -4.97 (0.07) & -4.82
(0.07) & -4.62 (0.04) & -4.89 (0.03)\\
$P_{\mathrm{rot}}(HK)^a$ & [days] & 39 & 51 & 42 & 22 & 12 & 14\\
Age &[Gy]                   & 4.3 (4) &4.5 (4)&7.4 (4.5)&2.4 (1.8)&1.3 (1)&4.3
(1)\\
$R_*$ & [R$_{\odot}$]  & 0.69 &0.79  & 1.04 & 1.16 & 1.14  & 1.34\\
\hline
\end{tabular}
\end{table*}

\begin{figure}[h]
\epsfig{file=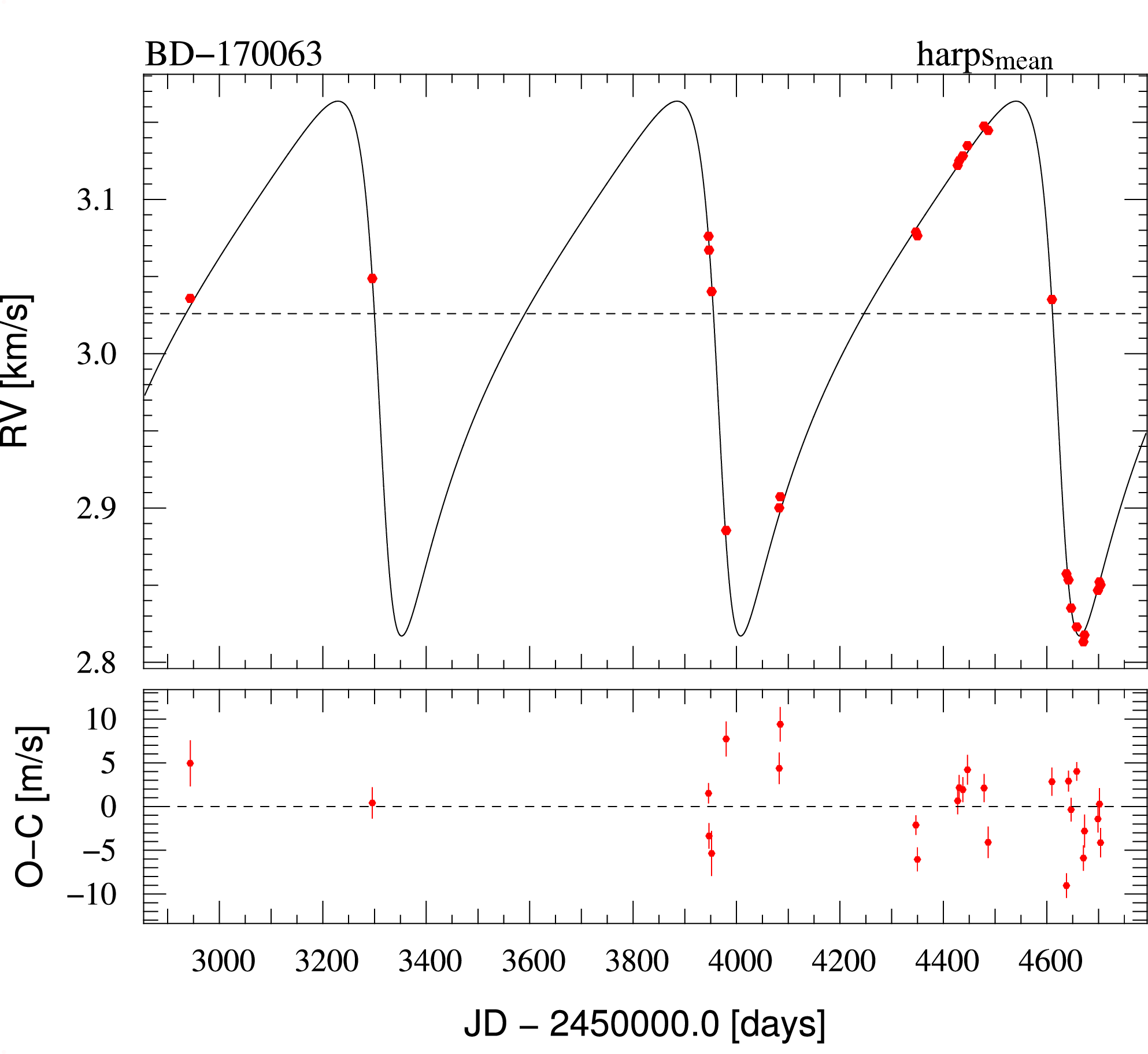,width=\linewidth}
\epsfig{file=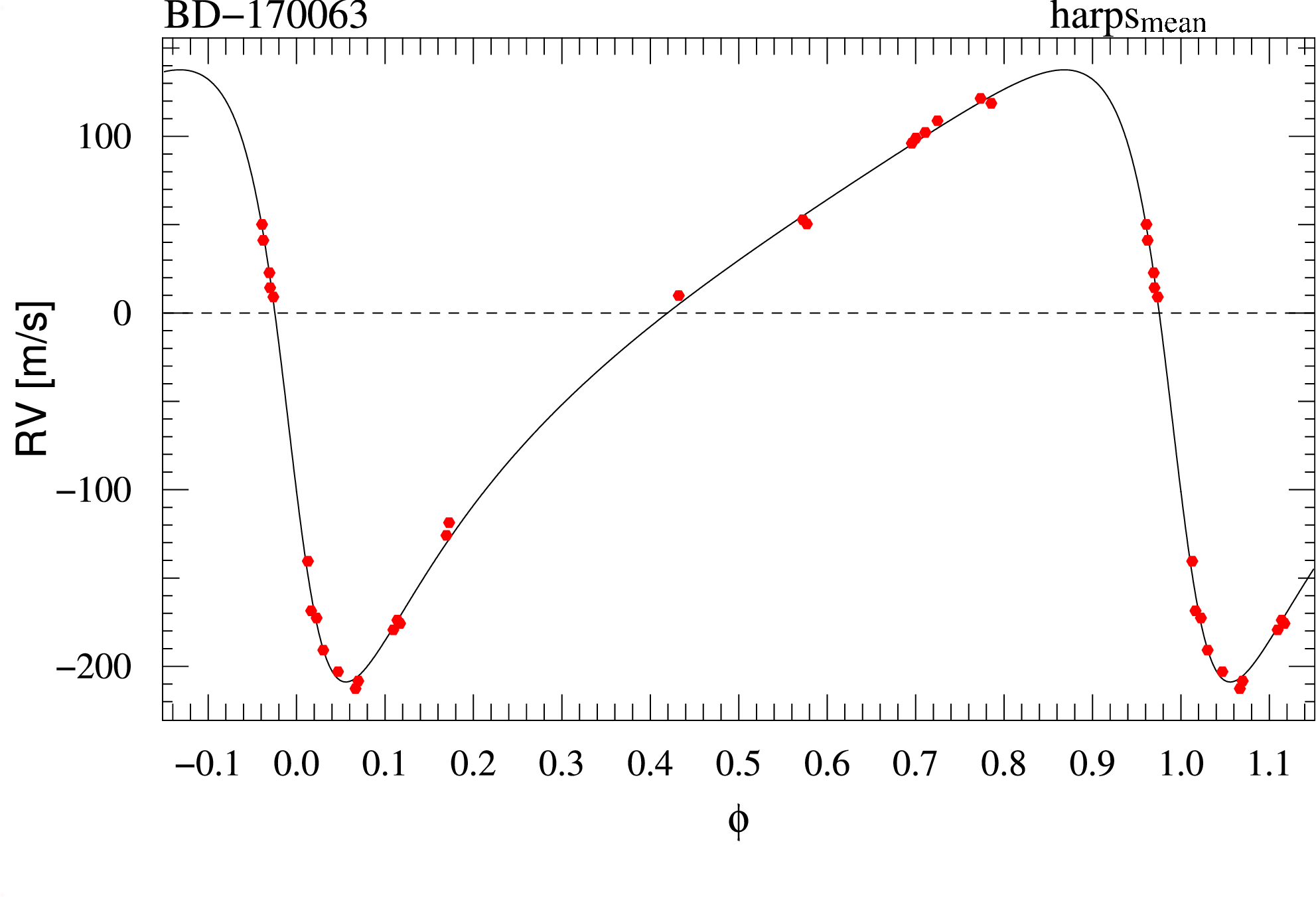,width=\linewidth}
\caption{The radial-velocity curve of BD\,-17 0063 obtained with HARPS. Top: 
  individual radial-velocity measurements (dots) versus time, and fitted
  orbital solution (solid curve); Middle: residuals to the fitted orbit versus
  time; Bottom: radial-velocity measurements with phase-folding, using the period of 655.6
  days and other orbital parameters as listed in Table 2. A 5.1 \mjup\,
  companion to this K5 dwarf is evidenced.}
\label{obs1}
\end{figure}

\section{Radial velocity data and orbital solutions}
\label{planet}
\subsection{BD\,-17 0063}
We gathered 26 spectra of BD\,-17 0063 with HARPS over a timespan of 1760 days between October
31st 2003 and July 5th 2008. The mean radial-velocity uncertainty is
1.6\,\ms. The measurements are given in Table \ref{rv1} (electronic version
only) and shown in Figure \ref{obs1}. We fitted a Keplerian orbit to the
observed radial-velocity variations, and found a best solution with a period
of 655.6 days. It is an eccentric orbit ($e=0.54$) with a semi-amplitude of
173 \ms. The reduced $\chi^2$ obtained on this fit is 3.2.

The inverse bisector slope is estimated on the cross-correlation function and
its timeseries is also examined, in order to exclude the stellar variability
as origin of the observed radial-velocity variation \citep{queloz2001}. The error on the bisector
slope is taken as twice the error on the velocity, as a conservative value. 
No correlation is found between the bisector slope and the velocity,
which excludes a blend scenario.  The bisector values 
for BD\,-17 0063 are compatible with a constant value with a
standard deviation of 7 \ms, over the 4.8 yr time span.
With the long rotation period
estimated for the star (39 days), a radial-velocity modulation related to spot
activity is also very unprobable.
These activity indicators thus strongly support the planetary origin of the
observed signal.

With the stellar parameters as determined in the previous section, we infer a
minimum planetary mass of $m_2 \sin{i}=$ 5.1 \mjup\, and semi-major axis of 1.34 AU. The periastron
distance is 0.87 AU which infers a transit probability of only 0.4\%. No
attempt was made yet to monitor the photometric lightcurve of BD\,-17 0063 nor to
search for a potential transit. Figure \ref{obs1} shows the radial-velocity signal
folded with the planetary phase and the residuals against time when the main
signal is subtracted. There is no significant periodic trend nor linear drift
in the O-C residuals, with a standard deviation of 4.1 \ms, i.e., marginally
above the individual errors.
All parameters of the orbit and the planet are given in Table \ref{TablePlanets},
together with their estimated error.

\begin{figure}[h]
\epsfig{file=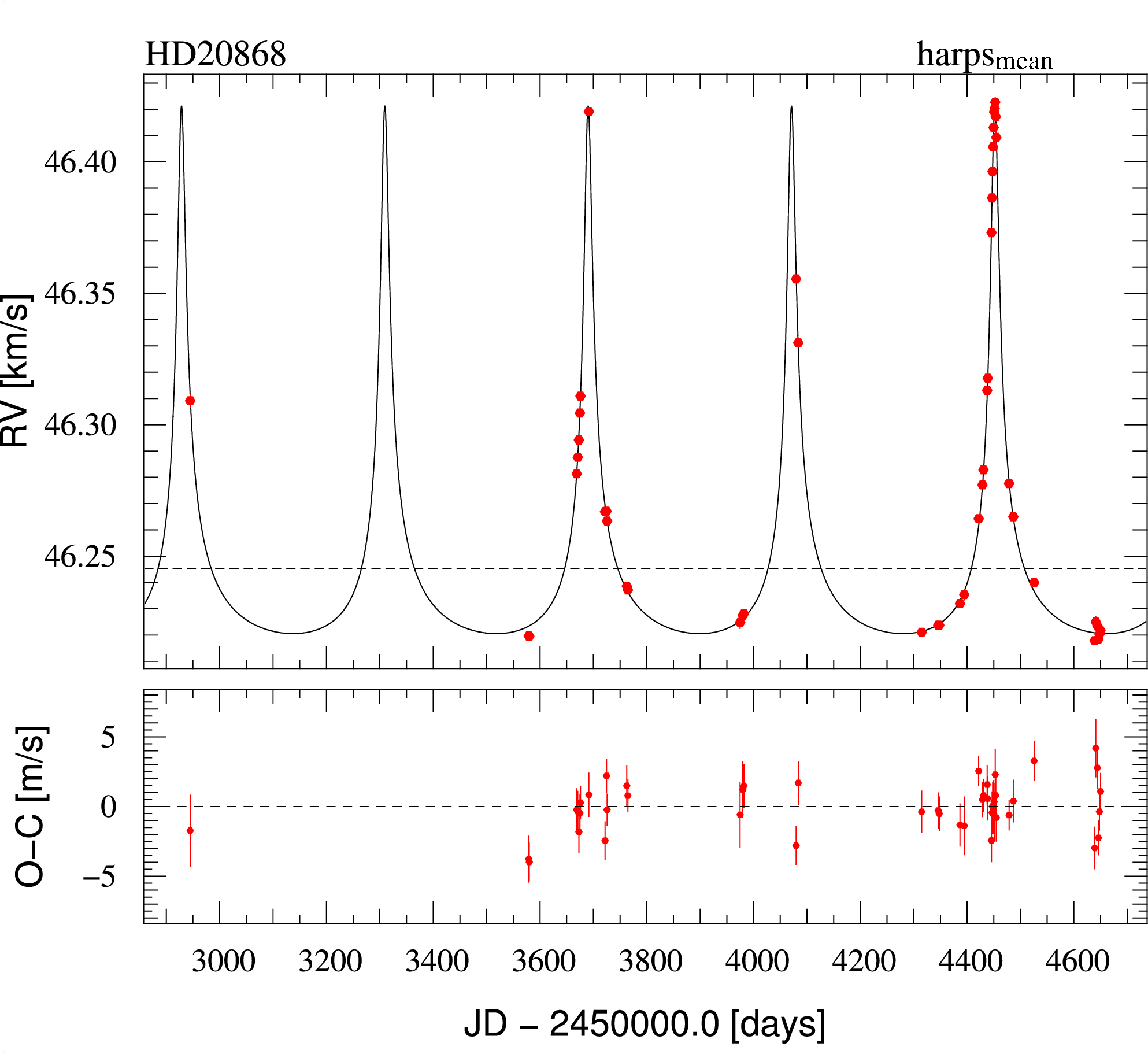,width=\linewidth}
\epsfig{file=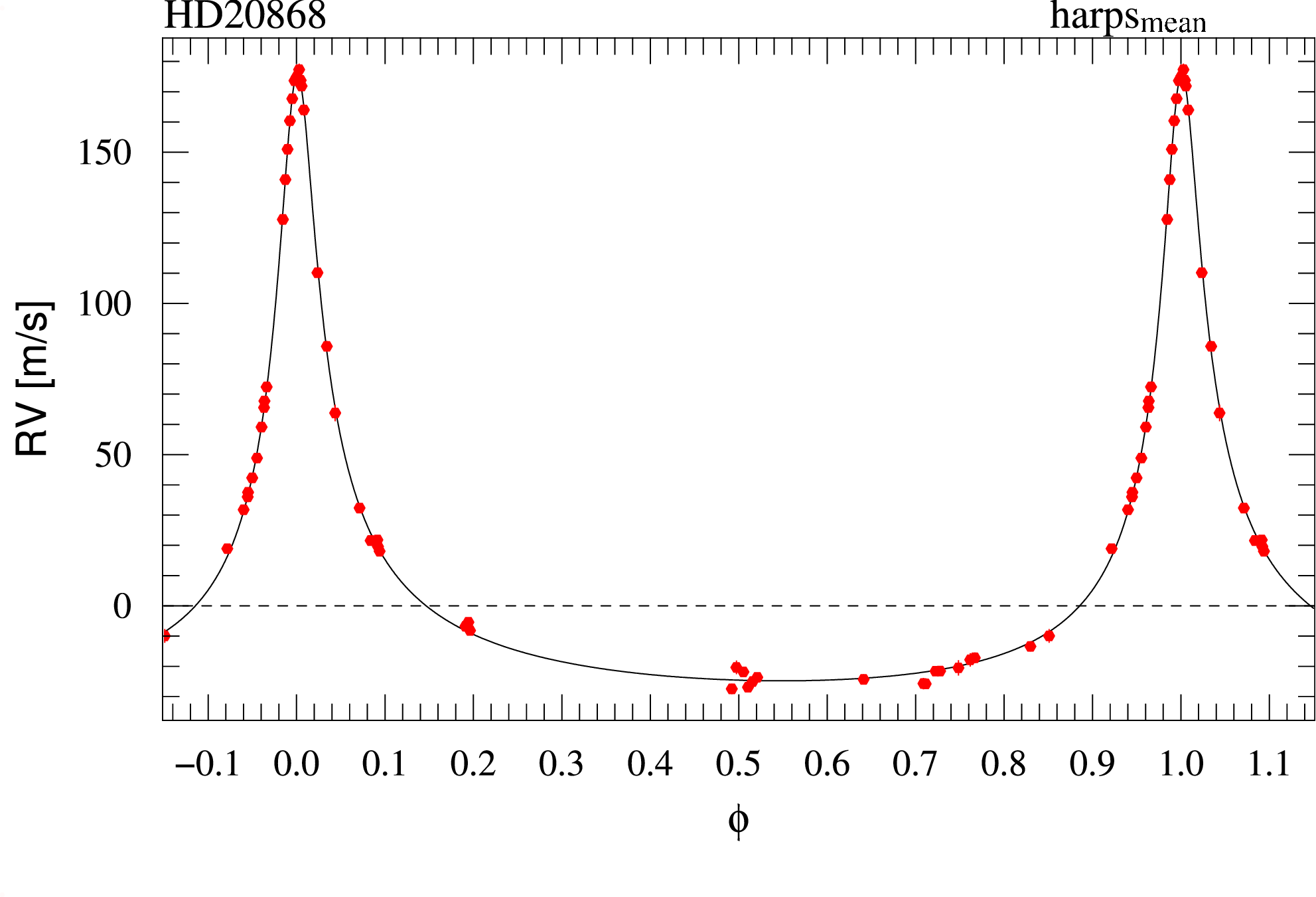,width=\linewidth}
\caption{The radial-velocity curve of HD\,20868 obtained with HARPS. Top: 
  individual radial-velocity measurements (dots) versus time, and fitted
  orbital solution (solid curve); Middle: residuals to the fitted orbit versus
  time; Bottom: radial-velocity measurements with phase-folding, using the period of 380.85
  days and other orbital parameters as listed in Table 2. The K3/4 IV star has
  a 1.99 \mjup\, companion.
}
\label{obs2}
\end{figure}

\subsection{HD\,20868}

Observations of HD\,20868 consist in  48 HARPS measurements obtained
over 1705 days between November 1st 2003 and July 2nd 2008. The mean
uncertainty on the radial velocity measurements is 1.5 \ms. The
measurements are given in Table \ref{rv2} (electronic version only). Figure \ref{obs2} shows the
velocities as
a function of time, as well as the Keplerian orbit with a period of 380.85 days
that best fits the data. The residual values, after subtraction of the fit,
are also shown against time. There is no significant trend in
these residuals, characterized by a standard deviation of 1.7 \ms. The reduced
$\chi^2$ of the fit is 1.27.

The best orbital solution is a strongly eccentric orbit ($e=$ 0.75) with a
semi-amplitude of 100.34 \ms. The inferred minimum mass of the companion responsible for
this velocity variation is 1.99 \mjup\, and a semi-major axis of 0.947 AU is
derived from the third Kepler law. The periastron distance is 0.54 AU which
corresponds to a transit probability of 0.7\%. 

The bisector test was applied and excludes that the velocity variations
are due to stellar activity. A trend which is confirmed by the long rotation period. 

\begin{figure}[h]
\epsfig{file=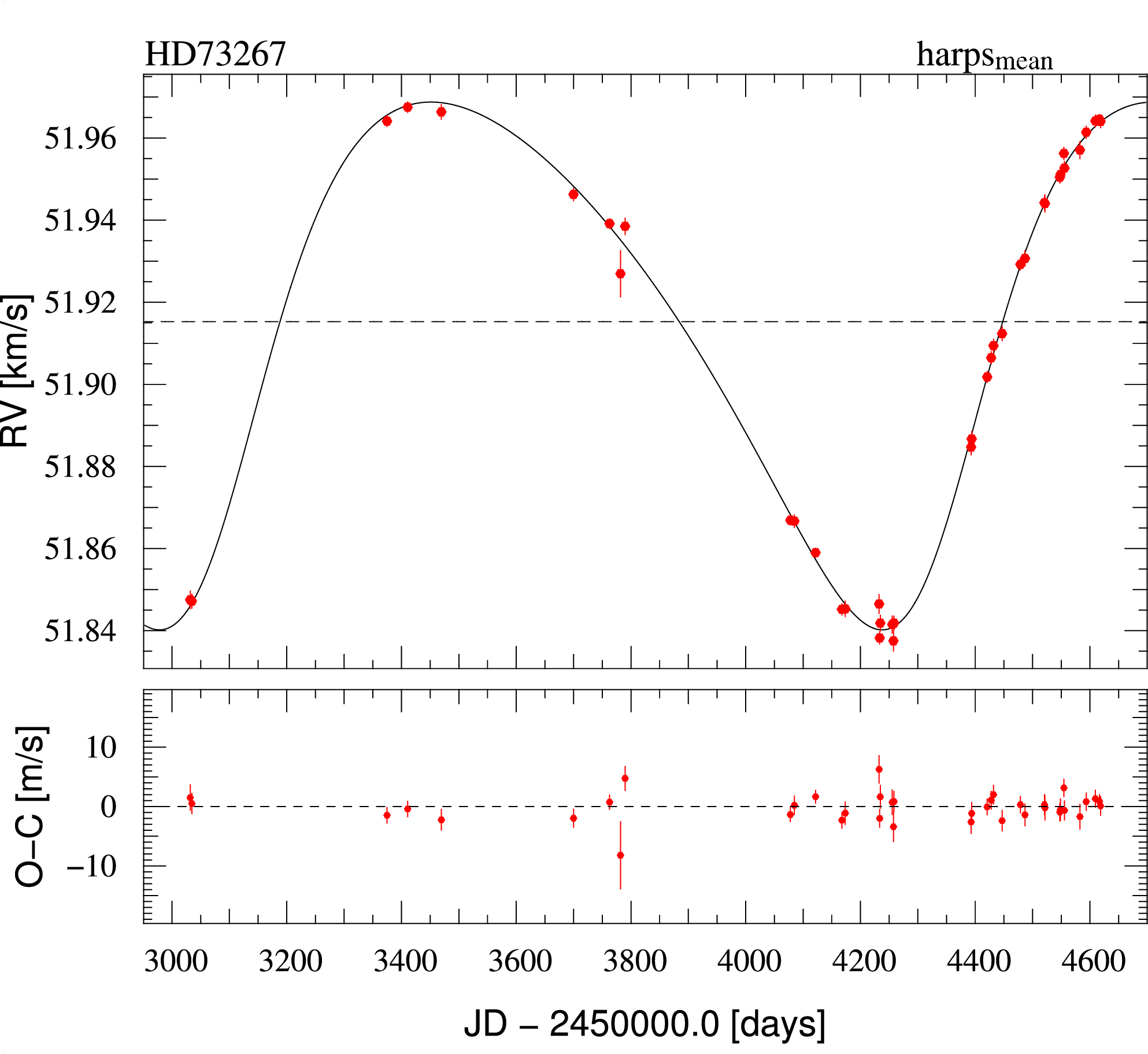,width=\linewidth}
\caption{The radial-velocity curve of HD\,73267 obtained with HARPS. 
  Top: individual radial-velocity measurements (dots) versus time, and fitted
  orbital solution (solid curve). The star has a companion of minimum mass
  3.06 \mjup\, and orbital period 1260 days. Bottom: residual to the fitted orbit versus time.}
\label{obs3}
\end{figure}

\subsection{HD\,73267}
We gathered 39 HARPS measurements of HD\,73267 over a time span of 1586
days, from November 27th 2004 and May 31st 2008. Small individual
uncertainties are obtained, with a mean value of 1.8 \ms. Data are shown in
Table \ref{rv3} and in Figure \ref{obs3}.
The observed velocity variations were fitted with a Keplerian orbit. The best solution
corresponds to a period of 1260 days, eccentricity of 0.256 and semi-amplitude
of 64.29 \ms. The scatter of the residuals is compatible with the
radial-velocity uncertainty and these residuals show no specific trend.
The $O-C$ standard deviation is 1.7 \ms\  and reduced $\chi^2$ is 1.19. 

The bisector variations are not correlated to the velocity
variations nor in phase with the signal, which excludes the stellar
variability as being the cause of it. Here again, the estimated rotation period of the
star is long, and spot-related activity cannot be considered as a
potential origin for the observed signal.

The minimum mass of the inferred companion is 3.06 \mjup\, and a semi-major axis of 2.198 AU
is calculated for this 3.44 year period companion.

\begin{figure}[h]
\epsfig{file=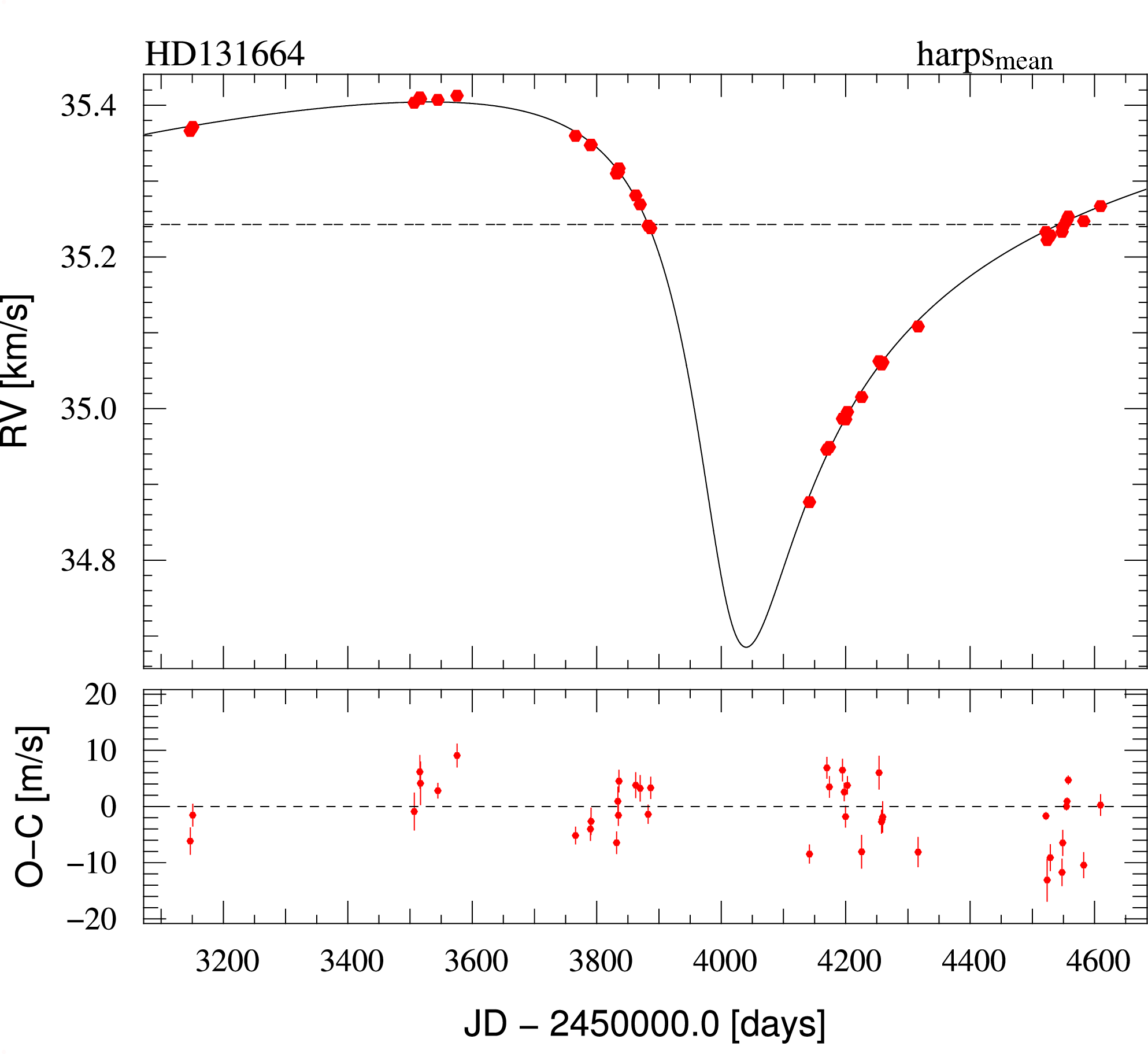,width=\linewidth}
\caption{The radial-velocity curve of HD\,131664 obtained with HARPS. 
  Top: individual radial-velocity measurements (dots) versus time, and fitted
  orbital solution (solid curve). It shows a very massive planetary companion
  of $m_2 \sin{i}=$18.15 \mjup\, with an orbital period of 1951 days. Bottom: 
  residual to the fitted orbit versus time.}
\label{obs4}
\end{figure}

\subsection{HD\,131664}
We gathered 41 measurements of HD\,131664 over 1463 days with HARPS,
from May 21st 2004 and May 23rd 2008. Individual uncertainties have a mean
value of 2 \ms. A long-term velocity variation is observed (Figure \ref{obs4}), which is best
fitted with a Keplerian orbit of 1951 days, 0.638 eccentricity and a large
semi-amplitude $K$ of 359.5 \ms. The residuals after subtraction of this signal have
a standard deviation of only 4 \ms\  and no specific trend. The reduced $\chi^2$
of the fit is 2.97. 
{\bf Although the orbital fit to the data appears robust, the time coverage of 
this planetary orbit is not perfect, as most of the periastron passage has unfortunately been missed. 
This limits the precision we get on the orbital parameters.  }

The bisector test
again confirms the origin of this signal as due to a sub-stellar
companion.
The large amplitude despite the long period of the signal results in a
large projected mass of the companion, i.e. $m_2 \sin{i}=$18.15 \mjup. This
massive planet, or brown-dwarf companion, orbits the parent
star at a semi-major axis of 3.17 AU.

\begin{figure}[h]
\epsfig{file=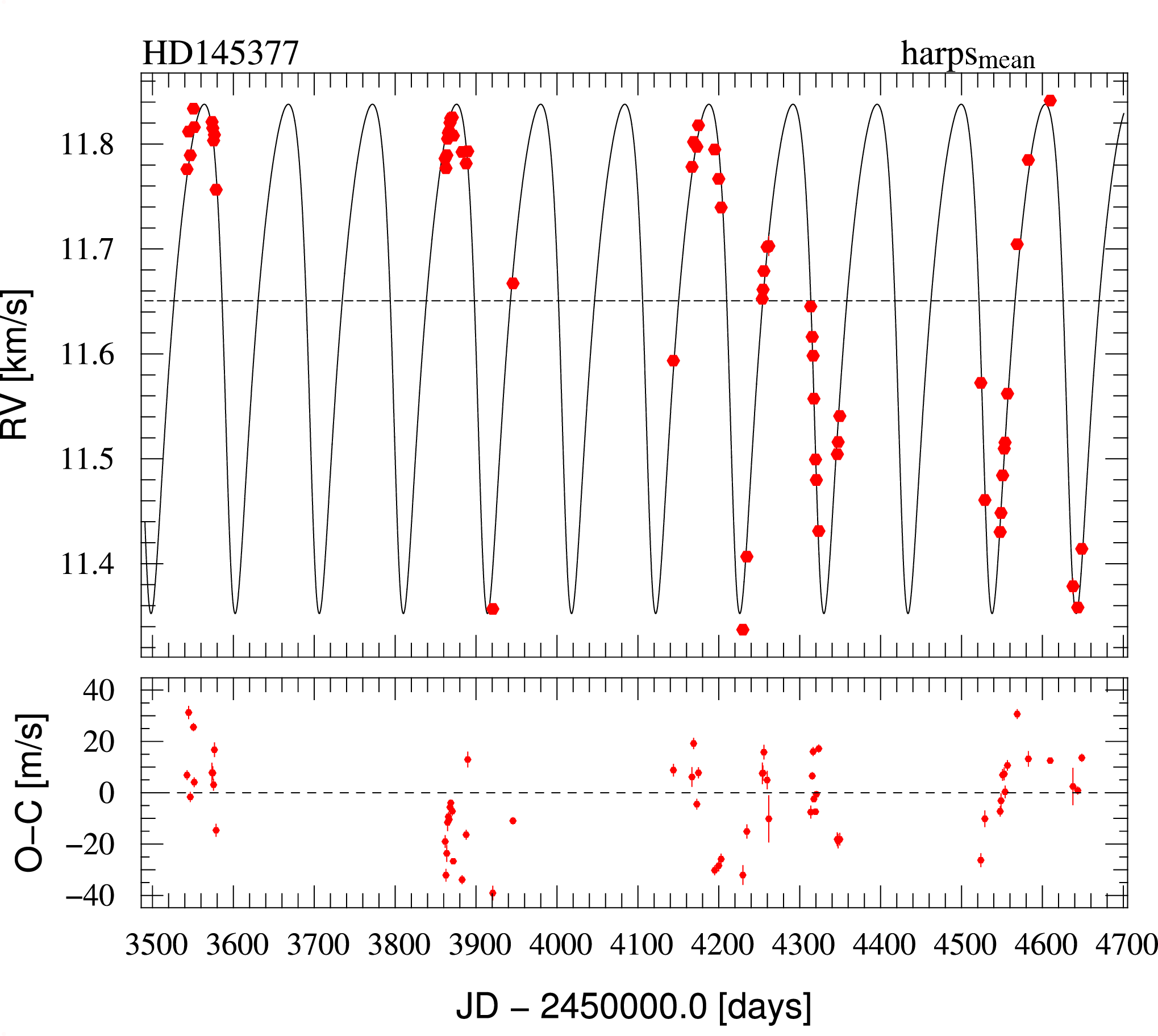,width=\linewidth}
\epsfig{file=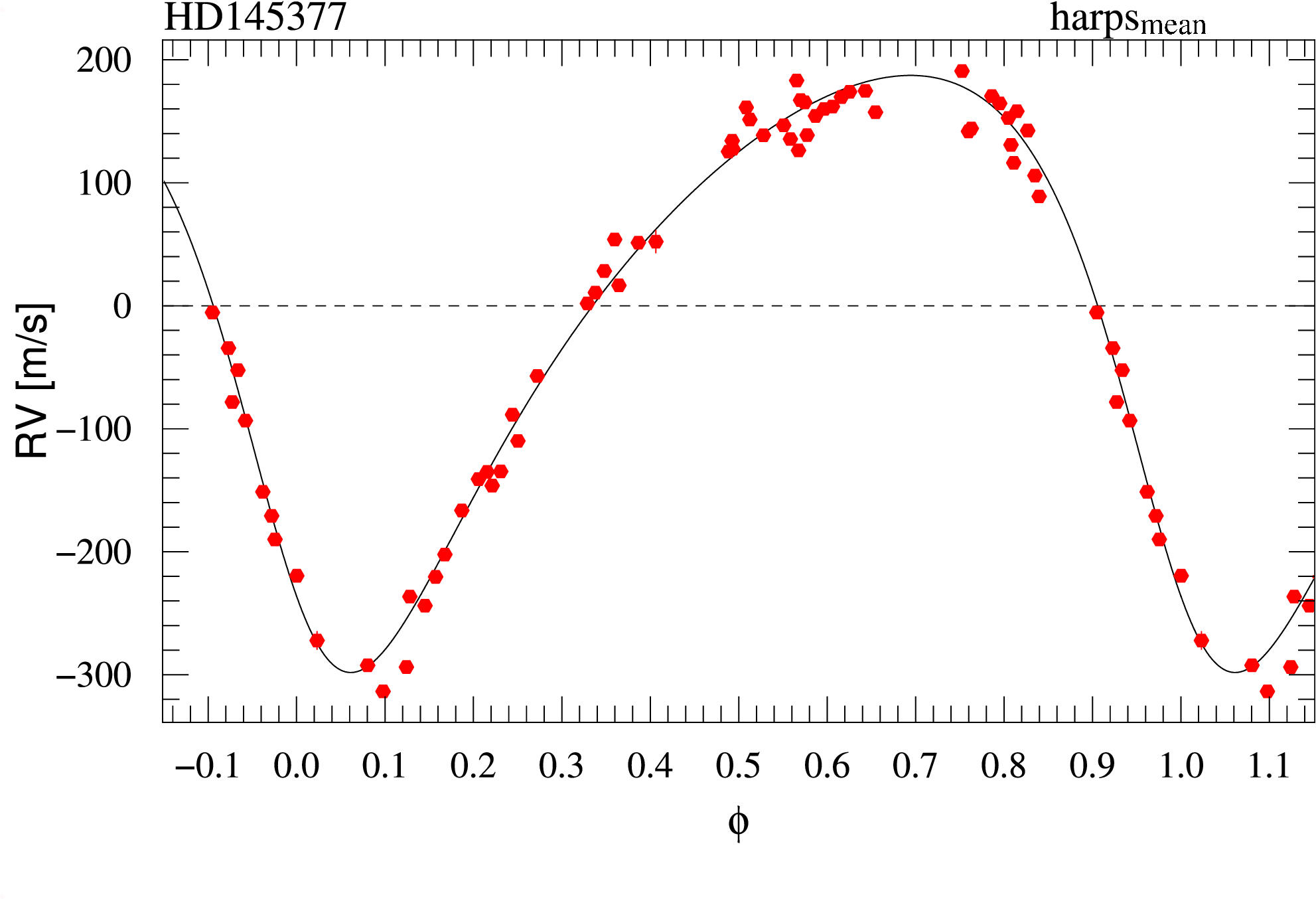,width=\linewidth}
\caption{The radial-velocity curve of HD\,145377 obtained with HARPS. Top: 
  individual radial-velocity measurements (dots) versus time, and fitted
  orbital solution (solid curve); Middle: residuals to the fitted orbit versus
  time; Bottom: radial-velocity measurements with phase-folding, using the period of 103.95
  days and other orbital parameters as listed in Table 2. The residual jitter
  is due to stellar variability (expected from activity indicators) and shows
  no periodic trend. A planet of minimum mass 5.76 \mjup\, is evidenced.}
\label{obs5}
\end{figure}

\begin{figure}[h!]
\epsfig{file=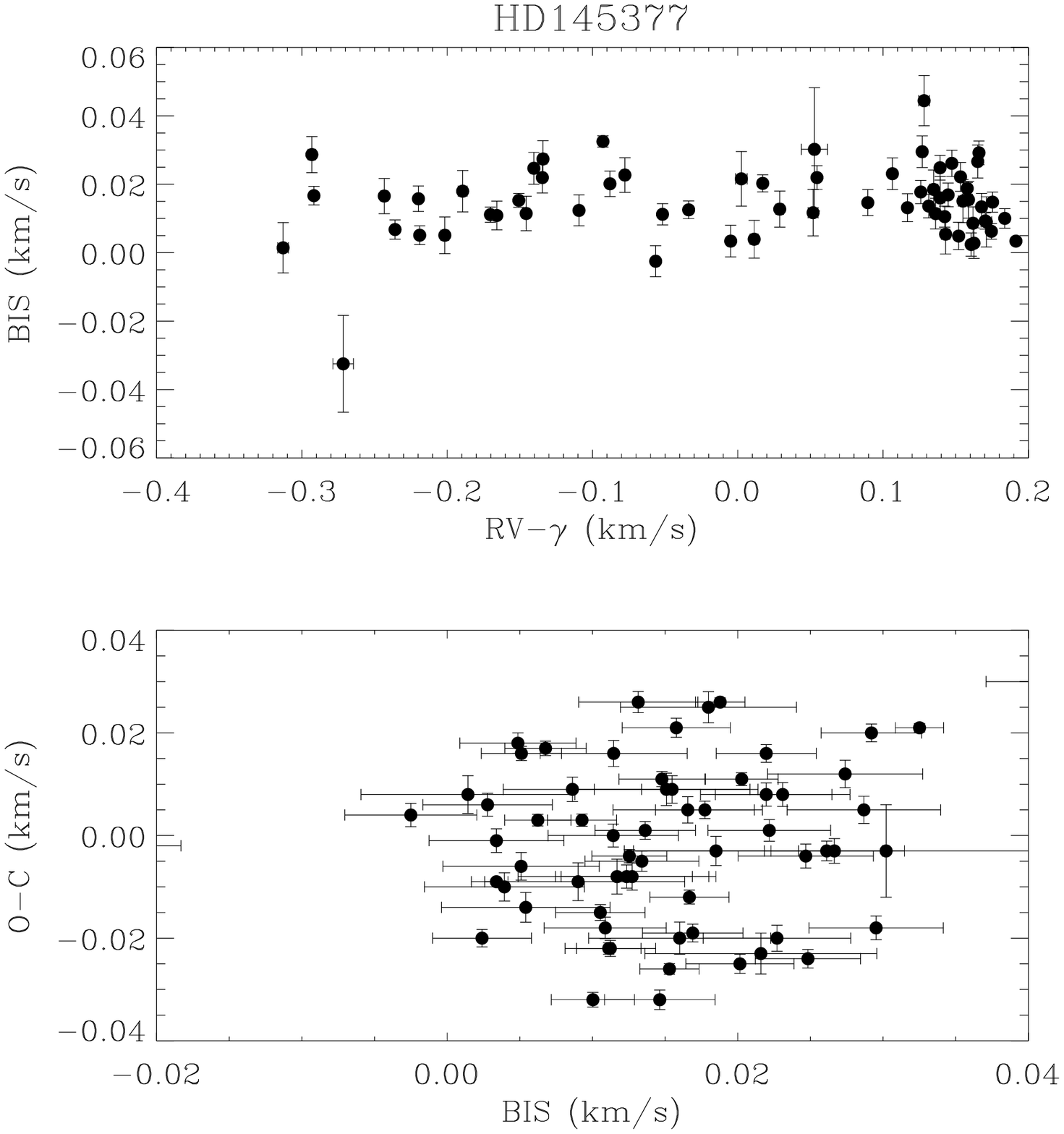,width=\linewidth}
\caption{The inverse bisector slope is plotted against the radial velocity of
  HD\,145377 (top) and against the residuals to the fitted orbit (bottom). No
  correlation between these quantities is observed. Although the bisector
  varies in a similar scale as the fit residuals, we cannot correct for
  spot-related activity. The amplitude of bisectors' variations still remains
  small compared to the range of radial velocities.}
\label{bis5}
\end{figure}

\subsection{HD\,145377}
We gathered 64 measurements of HD\,145377 with HARPS from June 21st
2005 to July 1st 2008, over 1106 days. A mean uncertainty of 2.3 \ms\  is
obtained. A relatively large-amplitude velocity variation is observed, as
shown in Figure \ref{obs5}. It is
best fitted with a 103.95 day period. The orbit is eccentric ($e=$ 0.307) and
semi-amplitude $K=$242.7 \ms. Although the signal is very clear and stable
over more than 10 periods, the residuals to the fit are affected by an
additional jitter, of
amplitude 15.3 \ms. This jitter was expected from the relatively young age (1
Gyr) and the high value of
log$R'_{HK}$ (mean value is -4.68),
which strongly suggests that stellar variability is observed in addition to
the main signal. The O-C residuals do not show, however, a periodicity {\bf related to the 12d rotation}, which is not surprising over about 80
rotation cycles of the star. {\bf The Lomb-Scargle periodogram of the residuals do show, however, a tendency for a curved drift that could be a hint for a second, longer-period planet, and other periodic signals could be present but too weak to be significant}. When taken into account, the curved drift decreases the residual noise from 15 to about 10 \ms. {\bf More data in the future may therefore reveal more planets in this system, but the present material is not conclusive in this respect}.

Figure \ref{bis5} shows the bisector behaviour with respect to radial
velocity (top) and as a function of the fit residuals. 
The scatter of the bisector span is larger than for the other stars,
with a value of 11 \ms, and it confirms that we see some line profile
variations with time. The bisector slope does not correlate, however, with the radial
velocity, excluding the stellar variability to be the only origin of the
observed velocity variation. 
We also find no correlation between the residuals to the
fitted orbit and the bisector span (Figure \ref{bis5} bottom). 
Such a correlation, observed when the activity is mainly related to spots, could have been used to correct
the radial velocities for stellar variability, as explained in \citet{melo07}.
Finally, as a test to the origin of the RV jitter, we observed HD\, 145377 in
a sequence of 10 consecutive 90s exposures, for which a standard deviation
of 2.5 \ms\ is derived. The stellar jitter therefore does not come from
short-term acoustic modes but rather from chromospheric activity features.

The planetary companion of HD\,145377 is a $m_2 \sin{i}=$5.76 \mjup\, planet orbiting with a
103.95d period. The semi-major axis is 0.45 AU. The periastron distance is
0.34 AU which corresponds to a transit probability of 0.14\%.

\subsection{HD\,153950}
\begin{figure}
\epsfig{file=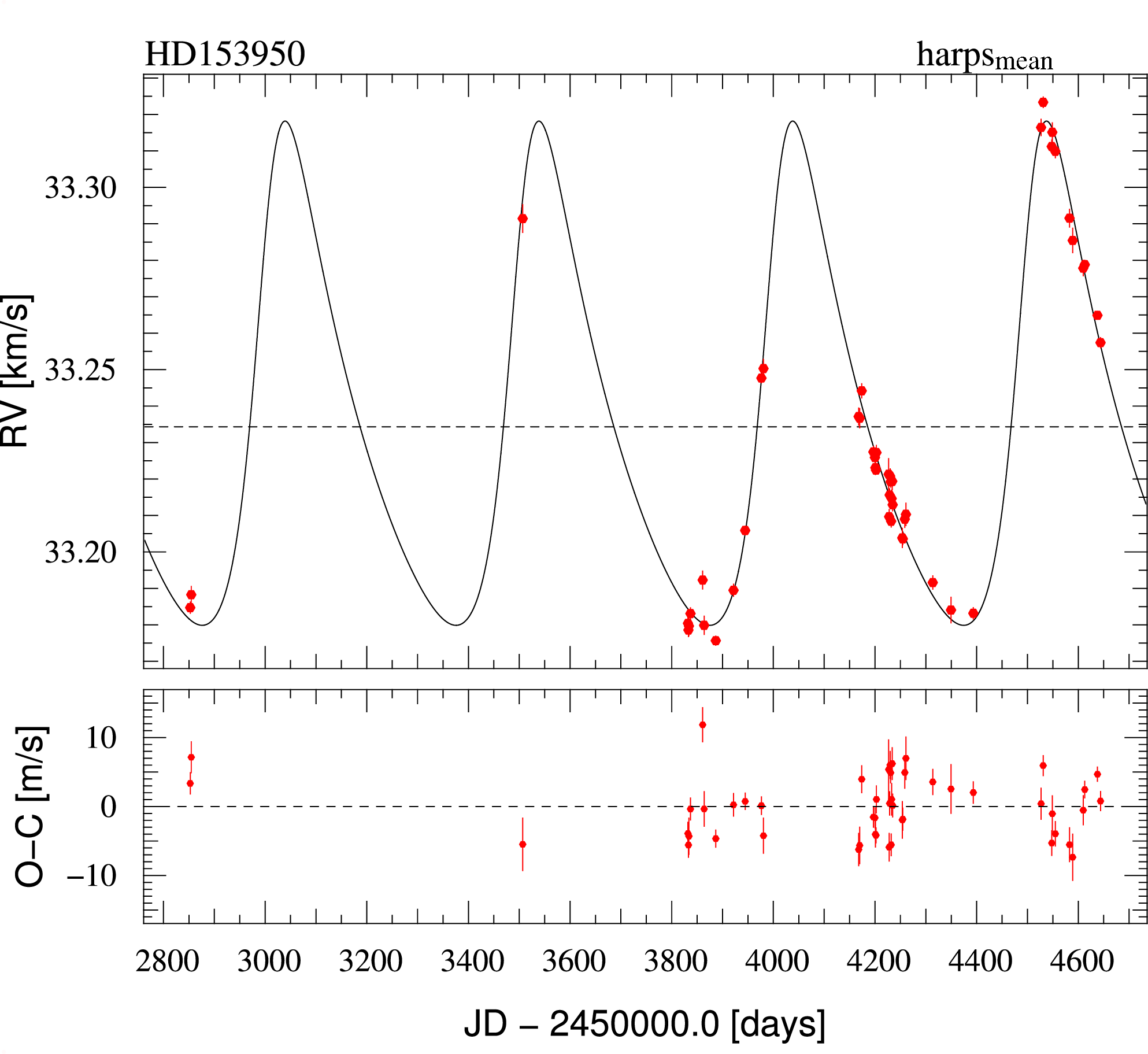,width=\linewidth}
\epsfig{file=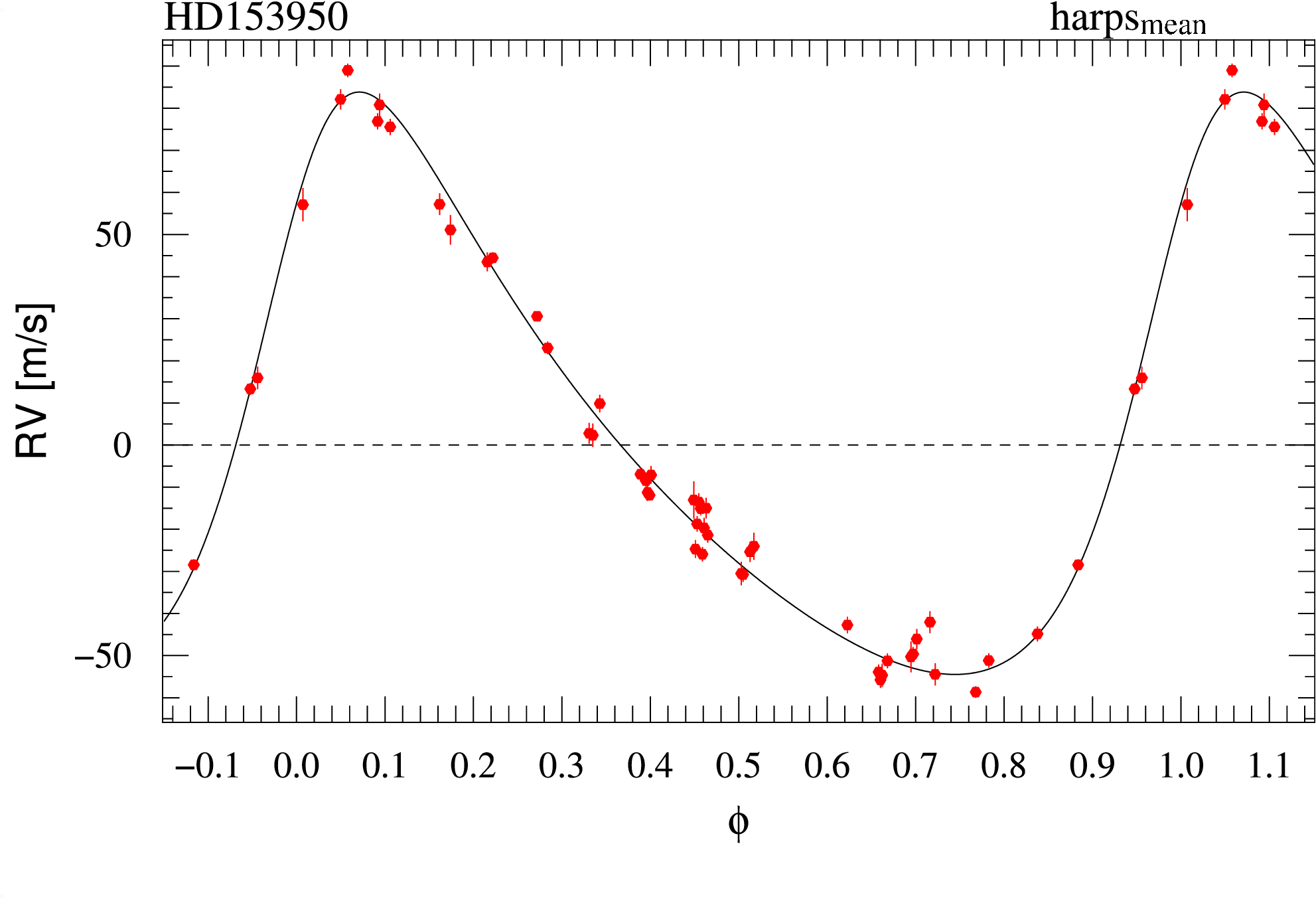,width=\linewidth}
\caption{The radial-velocity curve of HD\,153950 obtained with HARPS. Top: 
  individual radial-velocity measurements (dots) versus time, and fitted
  orbital solution (solid curve); Middle: residuals to the fitted orbit versus
  time; Bottom: radial-velocity measurements with phase-folding, using the period of 499.4
  days and other orbital parameters as listed in Table 2. The planetary
  companion has a minimum mass of 2.73 \mjup.}
\label{obs6}
\end{figure}

Finally, the star HD\,153950 was observed 49 times with HARPS from
August 1st 2003 to June 26th 2008 (1791 day time span). The mean uncertainty of
the velocity measurements is 2 \ms. The velocity variation with time is fitted with a
Keplerian orbit of 499.4 day period (Figure \ref{obs6}). It is again an eccentric orbit with
$e=$0.34 and a semi-amplitude of 69.15 \ms. The bisector is rather flat over
time and does not correlate with the orbital phase nor the position of the
velocity peak. The residuals around the best solution have a standard deviation of 4 \ms\ and
the reduced $\chi^2$ obtained for the fit is 2.40.

This radial-velocity curve thus shows a planetary companion of minimum mass 2.73 \mjup. Its
semi-major axis is 1.28 AU.\\

The orbit and planetary parameters of the six new systems described above are
given with their inferred errors in Table \ref{TablePlanets}.

\begin{table*} 
\caption{Orbital and physical parameters for the planets presented in
this paper. $T$ is the epoch of periastron. $\sigma$(O-C) is the residual
noise after orbital fitting of the combined set of measurements. $\chi^2_{red}$ is
the reduced $\chi^2$ of the fit.}
\label{TablePlanets}
\centering
\begin{tabular}{l l c c c c c c}
\hline\hline
\multicolumn{2}{l}{\bf Parameter}
& \bf BD\,-17 0063\,b & \bf HD\,20868\,b & \bf HD\,73267\,b & \bf HD\,131664\,b
& \bf HD\,145377\,b & \bf HD\,153950\,b \\
\hline
$P$ & [days] & 655.6 (0.6) 
             & 380.85 (0.09) 
             & 1260. (7) 
             & 1951. (41)
             & 103.95 (0.13)
             & 499.4 (3.6) \\
$T$ & [JD-2400000] & 54627.1 (1.5)  
                   & 54451.52 (0.1) 
                   & 51821.7 (16)  
                   & 52060. (41)
                   & 54635.4 (0.6)
                   & 54502. (4.3)\\
  $e$ &            & 0.54 (0.005) 
                   & 0.75 (0.002)  
                   & 0.256 (0.009)  
                   &0.638 (0.02)
                   & 0.307 (0.017) 
                   & 0.34 (0.021)\\
$\gamma$ & [km s$^{-1}$] & 3.026 (0.0012)  
                    & 46.245 (0.0003)
                    & 51.915 (0.0005) 
                    & 35.243 (0.004)
                    & 11.650 (0.003)
                    & 33.230 (0.001) \\
$\omega$ & [deg]    & 112.2 (1.9) 
                    & 356.2 (0.4) 
                    & 229.1 (1.8) 
                    & 149.7 (1.0)
                    & 138.1 (2.8)
                    & 308.2 (2.4) \\
$K$ & [m s$^{-1}$] & 173.3 (1.7) 
                   & 100.34 (0.42)
                   & 64.29 (0.48)
                   & 359.5 (22.3)
                   & 242.7 (4.6)
                   & 69.15 (1.2) \\
$a_1 \sin{i}$ & [10$^{-3}$ AU] & 8.76 &  2.305 & 7.196 & 49.678& 2.2074 & 2.981\\
$f(m)$ & [10$^{-9} $M$_{\odot}$] & 209.0 & 11.26 & 31.324 & 4295.665 & 132.788 & 14.174\\
$m_2 \sin{i}$ & [M$_{\mathrm{Jup}}$] & 5.1 (0.12)& 1.99 (0.05)  & 3.06 (0.07)
& 18.15 (0.35) & 5.76 (0.10) & 2.73 (0.05)\\
$a$ & [AU] & 1.34 (0.02) & 0.947 (0.012)& 2.198 (0.025)& 3.17 (0.03)& 0.45
(0.004)& 1.28 (0.01)\\
\hline
$N_{\mathrm{meas}}$ & & 26  & 48  & 39&41 &64 &49  \\
$Span$ & [days]       & 1760 & 1705 & 1586 &1463 & 1106& 1791 \\
$\sigma$ (O-C) & [m s$^{-1}$] & 4.1 & 1.7 & 1.7 &4.0 & 15.3& 3.9\\
$\chi^2_{red}$ & &3.2 & 1.27 & 1.19 & 2.97 & 10.13  & 2.40 \\
\hline
\end{tabular}
\end{table*}

\begin{figure}
\epsfig{file=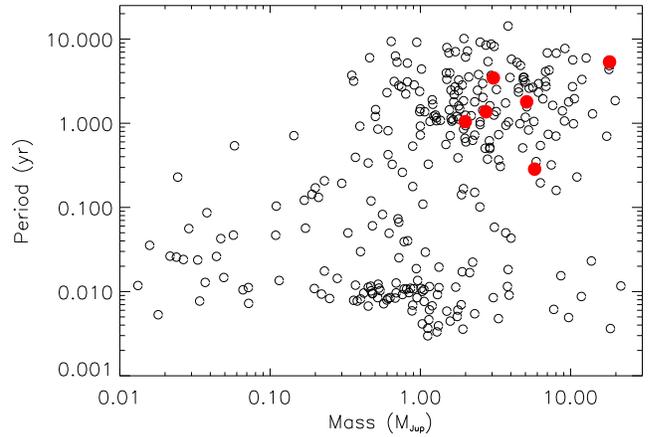,width=\linewidth}
\caption{The present mass-period diagram of known exoplanets (open circles)
  showing the location of the six new planets presented in this paper (filled
  circles). They belong to the bulge of the most massive, longest period bodies.}
\label{mp}
\end{figure}

\section{Conclusion}

From long-term observations with HARPS with individual uncertainty around 2\ms, we were able to derive the
presence of 6 new substellar companions around the main-sequence stars
BD\,-17 0063, HD\,20868, HD\,73267, HD\,131664, HD\,145377, and HD\,153950.

The analysis of the HARPS cross-correlation function and in particular the
bisector span of each measurement, allows to discard long-term stellar variability as
the origin of the observed radial-velocity curve. This proved efficient even
for the most active star, HD\,145377: characterizing the planetary companion 
does not suffer too much from the stellar variability, because of the planet's 
relatively short period with respect to our long time
span of observations (104 versus 1106 days). The stellar activity here
translates into a residual jitter that does not hide the planet's signal.

The planet orbiting HD\,131664 is very massive, with a minimum mass of 18.15 \mjup, over the
deuterium limit. It has characteristics similar to the other massive planet in
distant orbit HD\,168443\,c \citep{udry02}, except that no internal planet to
the system of HD\,131664 is evidenced so far. The period of HD\,131664\,b
(1951 days or 5.34 years) is
also among the dozen longest known so far.
Depending on the actual system age of HD\,131664, the magnitude
difference with the parent star could be as low as 13.5 in the
$K$ band -- for the lower edge of the age range -- and up to 20, 
using the models of \citet{baraffe02} for luminosity estimates. The angular
separation ranges from 0.035 to 0.16 arcsec during the orbit. Depending on the
system's inclination -- and thus the true mass of the companion -- it may be a
target for future direct imaging investigations, that would permit a better
characterisation of this unusual system. Note also that the parent star is
particularly metal-rich ($[Fe/H]=$0.32) in comparison with the mean metallicity of
the solar neighbourhood. The rare combination of parameters for this system --
companion's mass, orbital period, star's metallicity -- could make it one
important piece of constraints for theories of planetary formation.
Getting astrometric measurements of the six new systems with VLTI/PRIMA
would probably be possible, in order to better constrain their true mass.

The new planets discussed in this paper are part of the bulge of
long period, massive extrasolar planets in eccentric orbits, with masses in
the range 2-6 \mjup\, and periods of 0.3 to 5.3 years. Their properties can be
discussed in the frame of the statistical studies performed in 
well-defined stellar samples as in the ELODIE, CORALIE, or Lick+Keck+AAT
surveys (see \citet{marcy2005} and \citet{udry2007} for in-depth discussions):
\begin{itemize}
\item Giant gaseous planets are found around about 6-7\% of known main-sequence stars, with
  semi-major axes up to about 5 AU. These six new planets contribute to increase the
  number of known systems, with today 15 planets over the 850 stars of the
  volume-limited sample monitored by HARPS. The 1.8\% frequency of
  planet occurence in this sample where observations started in 2003, 
  is, however, not yet at the level of the oldest surveys. Identically, only
  five hot Jupiters were discovered in our sample, representing a frequency of
  0.6\%, to be compared to 1.2\% frequency for more complete surveys. The new
  planet sample presented in this paper yet contains the longest periods
  found in this specific survey, with measurements obtained at the earliest
  ages of HARPS operations (Figure \ref{mp}).
\item The distribution of planet masses currently favors the small masses,
  despite the strong observational bias towards massive planets.
  Here we bring
  new evidence for planets in the highest mass edge, with minimum masses of 2
  to 18 \mjup\ (Figure \ref{mp}). 
\item The period and eccentricity properties of the six new planets confirms
  the global tendency of very dispersed eccentricities beyond the
  circularization zone due to tidal interactions, compared to the circular
  orbits of giant and distant planets in the Solar System. 
  The origin of such dispersed eccentricities remains a
  mystery despite a number of theoretical attempts to match the observed
  distribution from a variety of eccentricity-damping physical processes.
\item Host stars of systems with giant gaseous planets 
  are significantly more metal-rich than average \citep{santos2005,fischer2005},
  which is not debated by the new exoplanet sample presented here,
  with two stars having an excess metallicity compared to the Sun and no
  metal-poor planet-host star. 
\item About 12\% of systems with gaseous giant planets are multiple. Here, we find no indication
for a second body in any of the new systems, with a very small scatter of the
residuals of the order of a few \ms (except for HD\,145377 which is active). 
In order to find planets of lower mass
in these systems, a high-precision strategy should now be applied. Finding
larger distance planets in these systems is also possible, although no
significant long-term drift is yet observed.
\item Finally, the mass-period distribution of the six new planets
  corroborates with the more general properties that more massive planets have
  longer orbital distances (e.g., \citet{udry2007}).  
\end{itemize}

Adding new extrasolar systems to the $\sim$300 planets known sofar is of course of
great importance to better characterize their properties. Radial-velocity
survey, as well as transit-search programs, suffer the observational bias of detecting more easily the short-period
and massive planets -the rarest ones-, which may be the reason why only 6-7\% of planets in the
solar neighbourhood show the signature of a giant planet. 
Note that this proportion of stars with planetary
systems greatly increases when planets in the mass range of Neptune or below
are discovered \citep{may2008}.
Extending the planet sample,
especially in well-defined volume-limited samples of
main-sequence stars as monitored by HARPS,
is one of the new challenges of
this scientific field, to help understanding the mechanisms which form and maintain
planets around other stars.

\onltab{3}{
\begin{table*}
\caption{Radial velocity values for BD\,-17 0063.}
\begin{tabular}{l c c}
\hline
JD-2,400,000.  &  Radial Vel. & Uncertainty \\
         & [km s$^{-1}$]   &  [km s$^{-1}$] \\
\hline
52943.610509  &  3.03585  &  0.00258  \\
53295.753206  &  3.04876  &  0.00174  \\
53945.903440  &  3.07611  &  0.00111  \\
53946.823277  &  3.06712  &  0.00141  \\
53951.864680  &  3.04030  &  0.00252  \\
53979.874025  &  2.88551  &  0.00194  \\
54082.599768  &  2.90011  &  0.00174  \\
54084.574580  &  2.90735  &  0.00192  \\
54346.836046  &  3.07882  &  0.00107  \\
54349.792760  &  3.07641  &  0.00132  \\
54427.598625  &  3.12206  &  0.00149  \\
54430.603057  &  3.12502  &  0.00140  \\
54437.605063  &  3.12820  &  0.00137  \\
54446.614295  &  3.13480  &  0.00164  \\
54478.592373  &  3.14747  &  0.00155  \\
54486.529176  &  3.14468  &  0.00174  \\
54609.910593  &  3.03512  &  0.00155  \\
54637.911547  &  2.85740  &  0.00136  \\
54641.924906  &  2.85339  &  0.00115  \\
54646.887252  &  2.83517  &  0.00130  \\
54657.823636  &  2.82295 &  0.00101  \\
54670.790307  &  2.81336 &  0.00139  \\
54672.838582  &  2.81770 &  0.00183  \\
54698.807055  &  2.84664 &  0.00149  \\
54701.823517  &  2.85211 &  0.00177  \\
54703.858883  &  2.85025 &  0.00162  \\
\hline
\end{tabular}
\label{rv1}
\end{table*}
}

\onltab{4}{
\begin{table*}
\caption{Radial velocity values for HD\,20868.}
\begin{tabular}{l c c}
\hline
JD-2400000.  &  Radial Vel. & Uncertainty \\
         & [km s$^{-1}$]   &  [km s$^{-1}$] \\
\hline
52944.759761  &  46.30911  &  0.00255  \\
53578.941288  &  46.21963  &  0.00165  \\
53579.927242  &  46.21953  &  0.00135  \\
53668.773536  &  46.28134  &  0.00150  \\
53670.728503  &  46.28763  &  0.00145  \\
53672.797736  &  46.29419  &  0.00148  \\
53674.743782  &  46.30446  &  0.00135  \\
53675.819517  &  46.31087  &  0.00114  \\
53691.577515  &  46.41910  &  0.00155  \\
53721.706478  &  46.26692  &  0.00135  \\
53724.633943  &  46.26712  &  0.00118  \\
53725.585119  &  46.26338  &  0.00113  \\
53762.579524  &  46.23860  &  0.00145  \\
53764.539118  &  46.23716  &  0.00113  \\
53974.869681  &  46.22482  &  0.00232  \\
53979.923748  &  46.22749  &  0.00200  \\
53981.897310  &  46.22813  &  0.00155  \\
54079.673981  &  46.35546  &  0.00135  \\
54083.717877  &  46.33114  &  0.00154  \\
54314.874429  &  46.22101  &  0.00149  \\
54345.833998  &  46.22372  &  0.00126  \\
54347.832433  &  46.22375  &  0.00119  \\
54386.726662  &  46.23193  &  0.00151  \\
54394.752760  &  46.23539  &  0.00207  \\
54421.692913  &  46.26423  &  0.00102  \\
54428.736985  &  46.27712  &  0.00121  \\
54430.633523  &  46.28286  &  0.00112  \\
54437.676734  &  46.31306  &  0.00138  \\
54438.644277  &  46.31772  &  0.00154  \\
54445.632867  &  46.37310  &  0.00152  \\
54446.708327  &  46.38629  &  0.00149  \\
54447.647299  &  46.39631  &  0.00173  \\
54448.662135  &  46.40572  &  0.00191  \\
54449.673081  &  46.41303  &  0.00144  \\
54450.625557  &  46.41899  &  0.00131  \\
54451.657624  &  46.42038  &  0.00175  \\
54452.629288  &  46.42265  &  0.00178  \\
54453.668760  &  46.41719  &  0.00140  \\
54454.671103  &  46.40928  &  0.00169  \\
54478.640303  &  46.27767  &  0.00106  \\
54486.573877  &  46.26497  &  0.00149  \\
54525.518526  &  46.23992  &  0.00137  \\
54638.948171  &  46.21787  &  0.00149  \\
54640.934733  &  46.22498  &  0.00206  \\
54643.941030  &  46.22350  &  0.00151  \\
54645.930747  &  46.21843  &  0.00121  \\
54647.933071  &  46.22028  &  0.00133  \\
54649.925384  &  46.22170  &  0.00129  \\
\hline
\end{tabular}
\label{rv2}
\end{table*}
}

\onltab{5}{
\begin{table*}
\caption{Radial velocity values for HD\,73267.}
\begin{tabular}{l c c}
\hline
JD-2400000.  &  Radial Vel. & Uncertainty \\
         & [km s$^{-1}$]   &  [km s$^{-1}$] \\
\hline
53031.730610  &  51.84750  &  0.00214  \\
53034.625318  &  51.84718  &  0.00171  \\
53374.800416  &  51.96412  &  0.00131  \\
53410.705484  &  51.96753  &  0.00130  \\
53469.507660  &  51.96638  &  0.00175  \\
53699.856836  &  51.94628  &  0.00153  \\
53762.700039  &  51.93917  &  0.00120  \\
53781.723770  &  51.92696  &  0.00565  \\
53789.716242  &  51.93850  &  0.00202  \\
54077.853694  &  51.86690  &  0.00117  \\
54084.808437  &  51.86669  &  0.00154  \\
54121.750283  &  51.85900  &  0.00107  \\
54167.636520  &  51.84519  &  0.00134  \\
54173.606470  &  51.84531  &  0.00188  \\
54232.560360  &  51.84650  &  0.00231  \\
54233.545820  &  51.83822  &  0.00148  \\
54234.516188  &  51.84184  &  0.00191  \\
54255.451119  &  51.84149  &  0.00207  \\
54257.498753  &  51.83751  &  0.00248  \\
54258.452112  &  51.84183  &  0.00170  \\
54392.856053  &  51.88474  &  0.00192  \\
54393.861259  &  51.88672  &  0.00183  \\
54420.846468  &  51.90180  &  0.00134  \\
54427.833648  &  51.90648  &  0.00148  \\
54431.832778  &  51.90942  &  0.00157  \\
54446.853818  &  51.91236  &  0.00172  \\
54478.822609  &  51.92923  &  0.00138  \\
54486.729809  &  51.93068  &  0.00185  \\
54520.667295  &  51.94426  &  0.00167  \\
54521.641787  &  51.94405  &  0.00208  \\
54547.580289  &  51.95049  &  0.00144  \\
54548.606686  &  51.95113  &  0.00184  \\
54554.625472  &  51.95626  &  0.00146  \\
54555.617466  &  51.95268  &  0.00161  \\
54582.554316  &  51.95708  &  0.00209  \\
54593.487209  &  51.96142  &  0.00148  \\
54609.467804  &  51.96419  &  0.00142  \\
54616.504677  &  51.96460  &  0.00111  \\
54618.463621  &  51.96406  &  0.00157  \\
\hline
\end{tabular}
\label{rv3}
\end{table*}
}

\onltab{6}{
\begin{table*}
\caption{Radial velocity values for HD\,131664.}
\begin{tabular}{l c c}
\hline
JD-2400000.  &  Radial Vel. & Uncertainty \\
         & [km s$^{-1}$]   &  [km s$^{-1}$] \\
\hline
53146.689878  &  35.36621  &  0.00235  \\
53150.721130  &  35.37138  &  0.00194  \\
53506.661928  &  35.40323  &  0.00328  \\
53515.682762  &  35.41041  &  0.00291  \\
53516.706183  &  35.40838  &  0.00380  \\
53544.650439  &  35.40707  &  0.00129  \\
53575.529526  &  35.41260  &  0.00204  \\
53765.883247  &  35.35969  &  0.00148  \\
53789.862142  &  35.34722  &  0.00200  \\
53790.891695  &  35.34791  &  0.00234  \\
53831.878129  &  35.30995  &  0.00191  \\
53833.894708  &  35.31519  &  0.00251  \\
53834.793072  &  35.31172  &  0.00179  \\
53835.766752  &  35.31675  &  0.00192  \\
53862.689370  &  35.28105  &  0.00222  \\
53869.760171  &  35.26920  &  0.00226  \\
53882.633600  &  35.24120  &  0.00160  \\
53886.645454  &  35.23779  &  0.00189  \\
54141.891213  &  34.87671  &  0.00160  \\
54169.833067  &  34.94571  &  0.00186  \\
54173.842331  &  34.94933  &  0.00183  \\
54194.919509  &  34.98657  &  0.00194  \\
54197.764290  &  34.98699  &  0.00156  \\
54199.875525  &  34.98575  &  0.00186  \\
54202.767256  &  34.99557  &  0.00155  \\
54225.722932  &  35.01516  &  0.00291  \\
54253.647281  &  35.06246  &  0.00292  \\
54257.633280  &  35.05809  &  0.00203  \\
54258.634595  &  35.05959  &  0.00167  \\
54259.617251  &  35.06109  &  0.00270  \\
54316.466834  &  35.10819  &  0.00261  \\
54521.868941  &  35.23283  &  0.00061  \\
54523.843516  &  35.22223  &  0.00376  \\
54528.845576  &  35.22825  &  0.00232  \\
54547.737400  &  35.23304  &  0.00236  \\
54548.830829  &  35.23871  &  0.00222  \\
54554.803145  &  35.24746  &  0.00060  \\
54555.817248  &  35.24875  &  0.00060  \\
54557.770244  &  35.25325  &  0.00071  \\
54582.619078  &  35.24712  &  0.00222  \\
54609.613271  &  35.26701  &  0.00185  \\
\hline
\end{tabular}
\label{rv3}
\end{table*}
}

\onltab{7}{
\begin{table*}
\caption{Radial velocity values for HD\,145377.}
\begin{tabular}{l c c}
\hline
JD-2400000.  &  Radial Vel. & Uncertainty \\
         & [km s$^{-1}$]   &  [km s$^{-1}$] \\
\hline
53542.633832  &  11.77595  &  0.00170  \\
53544.730958  &  11.81187  &  0.00238  \\
53546.751703  &  11.78922  &  0.00181  \\
53550.654537  &  11.83375  &  0.00143  \\
53551.609534  &  11.81601  &  0.00173  \\
53573.602135  &  11.82109  &  0.00367  \\
53574.531691  &  11.81516  &  0.00241  \\
53575.546089  &  11.80329  &  0.00211  \\
53576.546067  &  11.80878  &  0.00268  \\
53578.665360  &  11.75643  &  0.00232  \\
53861.783012  &  11.78616  &  0.00224  \\
53862.741865  &  11.77693  &  0.00231  \\
53863.749757  &  11.78931  &  0.00313  \\
53864.742775  &  11.80499  &  0.00315  \\
53865.771405  &  11.81063  &  0.00170  \\
53866.724997  &  11.81256  &  0.00223  \\
53867.778590  &  11.82035  &  0.00119  \\
53868.748140  &  11.82455  &  0.00114  \\
53870.593760  &  11.82535  &  0.00148  \\
53871.786449  &  11.80792  &  0.00085  \\
53882.713288  &  11.79242  &  0.00154  \\
53887.717218  &  11.78144  &  0.00173  \\
53889.679215  &  11.79304  &  0.00290  \\
53920.603556  &  11.35688  &  0.00264  \\
53945.583464  &  11.66717  &  0.00125  \\
54143.897228  &  11.59350  &  0.00227  \\
54166.891741  &  11.77823  &  0.00367  \\
54168.883742  &  11.80202  &  0.00200  \\
54172.857726  &  11.79734  &  0.00192  \\
54174.848564  &  11.81779  &  0.00196  \\
54194.917474  &  11.79477  &  0.00173  \\
54199.899854  &  11.76678  &  0.00205  \\
54202.889292  &  11.73950  &  0.00190  \\
54229.719267  &  11.33710  &  0.00368  \\
54234.652538  &  11.40673  &  0.00257  \\
54253.744628  &  11.65247  &  0.00399  \\
54254.678126  &  11.66130  &  0.00275  \\
54255.722560  &  11.67890  &  0.00264  \\
54259.725812  &  11.70184  &  0.00340  \\
54261.794683  &  11.70275  &  0.00901  \\
54313.620619  &  11.64514  &  0.00232  \\
54315.497608  &  11.61620  &  0.00129  \\
54316.611584  &  11.59821  &  0.00156  \\
54317.496297  &  11.55722  &  0.00083  \\
54319.542744  &  11.49931  &  0.00102  \\
54320.580289  &  11.47979  &  0.00111  \\
54323.537296  &  11.43111  &  0.00138  \\
54346.502395  &  11.50437  &  0.00253  \\
54347.501198  &  11.51593  &  0.00267  \\
54349.526984  &  11.54072  &  0.00226  \\
54523.859926  &  11.57233  &  0.00254  \\
54528.878747  &  11.46067  &  0.00303  \\
54547.759621  &  11.43022  &  0.00186  \\
54548.852149  &  11.44839  &  0.00269  \\
54550.842175  &  11.48418  &  0.00210  \\
54552.800759  &  11.50967  &  0.00233  \\
54553.821042  &  11.51551  &  0.00226  \\
54556.798027  &  11.56205  &  0.00186  \\
54568.811680  &  11.70450  &  0.00172  \\
54582.623731  &  11.78479  &  0.00284  \\
54609.642688  &  11.84143  &  0.00040  \\
54637.784021  &  11.37846  &  0.00708  \\
54643.725606  &  11.35829  &  0.00136  \\
54648.688288  &  11.41413  &  0.00140  \\
\hline
\end{tabular}
\label{rv3}
\end{table*}
}

\onltab{8}{
\begin{table*}
\caption{Radial velocity values for HD\,153950.}
\begin{tabular}{l c c}
\hline
JD-2400000.  &  Radial Vel. & Uncertainty \\
         & [km s$^{-1}$]   &  [km s$^{-1}$] \\
\hline
52852.541834  &  33.18472  &  0.00154  \\
52854.596492  &  33.18828  &  0.00225  \\
53506.793032  &  33.29144  &  0.00381  \\
53831.893340  &  33.18043  &  0.00164  \\
53832.896004  &  33.17858  &  0.00178  \\
53833.897869  &  33.17967  &  0.00263  \\
53836.894963  &  33.18310  &  0.00160  \\
53860.919573  &  33.19229  &  0.00247  \\
53863.788490  &  33.17988  &  0.00251  \\
53886.726351  &  33.17567  &  0.00124  \\
53921.606991  &  33.18949  &  0.00163  \\
53944.556842  &  33.20589  &  0.00119  \\
53976.479331  &  33.24768  &  0.00128  \\
53980.560756  &  33.25030  &  0.00255  \\
54167.860472  &  33.23714  &  0.00234  \\
54169.858171  &  33.23668  &  0.00264  \\
54173.850015  &  33.24419  &  0.00195  \\
54196.826071  &  33.22742  &  0.00148  \\
54199.882018  &  33.22591  &  0.00144  \\
54200.808656  &  33.22309  &  0.00185  \\
54201.910735  &  33.22244  &  0.00109  \\
54202.795181  &  33.22726  &  0.00197  \\
54226.886058  &  33.22130  &  0.00430  \\
54227.810261  &  33.20965  &  0.00199  \\
54228.798529  &  33.21562  &  0.00169  \\
54229.766644  &  33.22080  &  0.00196  \\
54230.838756  &  33.21925  &  0.00146  \\
54231.794935  &  33.20842  &  0.00157  \\
54232.713425  &  33.21465  &  0.00229  \\
54233.903107  &  33.21937  &  0.00230  \\
54234.783932  &  33.21295  &  0.00165  \\
54253.698720  &  33.20385  &  0.00265  \\
54254.731349  &  33.20359  &  0.00161  \\
54258.659591  &  33.20899  &  0.00227  \\
54260.837471  &  33.21031  &  0.00308  \\
54313.649435  &  33.19161  &  0.00183  \\
54349.550815  &  33.18406  &  0.00353  \\
54393.493188  &  33.18317  &  0.00155  \\
54526.824653  &  33.31645  &  0.00225  \\
54530.902047  &  33.32334  &  0.00145  \\
54547.786305  &  33.31122  &  0.00178  \\
54548.903849  &  33.31512  &  0.00259  \\
54554.855275  &  33.30989  &  0.00178  \\
54582.757801  &  33.29154  &  0.00245  \\
54588.928823  &  33.28545  &  0.00336  \\
54609.734744  &  33.27785  &  0.00211  \\
54612.769428  &  33.27880  &  0.00121  \\
54637.793172  &  33.26491  &  0.00103  \\
54643.794707  &  33.25740  &  0.00139  \\
\hline
\end{tabular}
\label{rv3}
\end{table*}
}

\begin{acknowledgements}
N.C.S. would like to thank the support from Funda\c{c}\~ao para
a Ci\^encia e a Tecnologia, Portugal, through programme
Ci\^encia\,2007 (C2007-CAUP-FCT/136/2006).
We are grateful to the ESO staff for their support during observations.
\end{acknowledgements}

\bibliographystyle{aa}
\bibliography{references}

\end{document}